\pdfoutput=1
\RequirePackage{ifpdf}
\ifpdf % We are running pdfTeX in pdf mode
\documentclass[pdftex]{sigma}
\else
\documentclass{sigma}
\fi

\numberwithin{equation}{section}

\begin{document}

%\allowdisplaybreaks

\renewcommand{\PaperNumber}{069}

\FirstPageHeading

\ShortArticleName{Quasicomplex ${\cal N}=2$, $d=1$ Supersymmetric Sigma Models}

\ArticleName{Quasicomplex $\boldsymbol{{\cal N}=2}$, $\boldsymbol{d=1}$
Supersymmetric\\ Sigma Models}

\Author{Evgeny A.~IVANOV~$^\dag$ and Andrei V.~SMILGA~$^\ddag$}

\AuthorNameForHeading{E.A.~Ivanov and A.V.~Smilga}

\Address{$^\dag$~Bogoliubov Laboratory of Theoretical Physics, JINR, 141980 Dubna, Russia}
\EmailD{\href{mailto:eivanov@theor.jinr.ru}{eivanov@theor.jinr.ru}}

\Address{$^\ddag$~SUBATECH, Universit\'e de Nantes, 4 rue Alfred Kastler, BP 20722, Nantes 44307,
France\footnote{On leave of absence from ITEP, Moscow, Russia.}}
\EmailD{\href{mailto:smilga@subatech.in2p3.fr}{smilga@subatech.in2p3.fr}}

\ArticleDates{Received June 30, 2013, in f\/inal form November 13, 2013; Published online November 18, 2013}

\Abstract{We derive and discuss a~new type of ${\cal N}=2$ supersymmetric quantum mechanical sigma models
which appear when the superf\/ield action of the (${\bf 1},{\bf 2},{\bf 1}$) multiplets is modif\/ied by
adding an imaginary antisymmetric tensor to the target space metric, thus completing the latter to
a~non-symmetric Hermitian metric.
These models are not equivalent to the standard de Rham sigma models, but are related to them through
a~certain special similarity transformation of the supercharges.
On the other hand, they can be obtained by a~Hamiltonian reduction from the complex supersymmetric ${\cal
N}=2$ sigma models built on the multiplets (${\bf 2},{\bf 2},{\bf 0}$) and describing the Dolbeault complex
on the manifolds with proper isometries.
We study in detail the extremal two-dimensional case, when the target space metric is def\/ined solely by
the antisymmetric tensor, and show that the corresponding quantum systems reveal a~hidden ${\cal N}=4$
supersymmetry.}


\Classification{81Q60; 81T60; 14F40}

\section{Introduction}

Supersymmetric quantum mechanical (SQM) sigma models have been studied for more than 30 years.
The Lagrangian of the simplest and most known model of this type has the form~\cite{FT}
\begin{gather}\label{121stand}
L=-\frac12\int d\bar\theta d\theta\, g_{MN}(X)DX^M\bar D X^N,
\end{gather}
where $X^M$ are the ({\bf 1},{\bf 2},{\bf 1}) superf\/ields of one-dimensional ${\cal N}=2$
supersymmetry\footnote{We use the notation of~\cite{PT} where the numerals count the numbers of the dynamic
bosonic, dynamic fermionic, and auxiliary bosonic f\/ields.},
\begin{gather}\label{XM}
X^M=x^M+\theta\psi^M+\bar\psi^M\bar\theta+F^M\theta\bar\theta,
\end{gather}
and $D$, $\bar D$ are the $d=1$ superspace covariant derivatives,
\begin{gather*}
D=\frac\partial{\partial\theta}-i\bar\theta\frac{\partial}{dt},
\qquad
\bar D=-\frac{\partial}{\partial\bar\theta}+i\theta\frac{\partial}{dt}.
\end{gather*}

In components, this gives
\begin{gather*}
\begin{split}
& L = \frac12g_{MN}\left[\dot{x}^M\dot{x}^N+F^M F^N+i\big(\bar\psi^M\nabla\psi^N-\nabla\bar\psi^M\psi^N\big)\right]
\\
& \hphantom{L=}{}+\Gamma_{M,PQ}F^M\psi^P\bar\psi^Q+\frac12(\partial_M\partial_Q g_{PN}
)\bar\psi^M\psi^N\bar\psi^P\psi^Q.
\end{split}
\end{gather*}
Here $\Gamma_{M, PQ}$ are the Christof\/fel symbols and $\nabla \psi^N = \dot{\psi}^N +
\Gamma^N_{PQ}\dot{x}^P \psi^Q $.
After eliminating the auxiliary f\/ields $F^M$, the Lagrangian acquires the form
\begin{gather*}%\label{LbezF}
L=\frac12g_{MN}\big[\dot{x}^M\dot{x}^N+i\big(\bar\psi^M\nabla\psi^N-\nabla\bar\psi^M\psi^N\big)\big]+\frac12R_{MNPQ}
\bar\psi^M\psi^N\bar\psi^P\psi^Q,
\end{gather*}
where $R_{MNPQ}$ is the Riemann tensor.

The geometrical interpretation of this model is also well known.
Its Hilbert space can be mapped onto the space of dif\/ferential forms such that supercharges can be
interpreted as the exterior derivative operator and its complex conjugate.
Thereby, the system amounts to the de Rham complex~\cite{Witgeom,Witgeom1}.

Another complex known to mathematicians, the Dolbeault complex which is def\/ined on a~complex manifold and
involves only holomorphic $(p,0)$-forms, can also be formulated as a~supersymmetric sigma model.
Its superf\/ield action~\cite{Hull,ISDir} involves ({\bf 2},{\bf 2},{\bf 0}) superf\/ields,
\begin{gather*}
Z^j=z^j+\sqrt{2}\theta\psi^j-i\theta\bar\theta\dot{z}^j,
\qquad
\bar Z^{\bar{j}}=\bar z^{\bar{j}}-\sqrt{2}\bar\theta\bar\psi^{\bar{j}}
+i\theta\bar\theta\dot{\bar{z}}^{\bar{j}},
\end{gather*}
$j, \bar j = 1,\ldots,n$, $\bar D Z = D \bar Z = 0$.

In the simplest case,
\begin{gather*}%\label{start}
S=-\frac{1}{4}\int dt d^2\theta\, h_{j\bar k}(Z,\bar Z)D Z^j\bar D\bar Z^{\bar k},
\end{gather*}
where $h_{j\bar k}$ is a~Hermitian complex metric.

Coming back to~\eqref{121stand}, it describes the de Rham complex provided the tensor $g_{MN}$ is symmetric
and has the meaning of the target space metric.
However, nothing prevents us from asserting that this tensor has also an imaginary antisymmetric part (we
should keep $g_{MN}$ Hermitian, otherwise we would lose the reality of the Lagrangian and Hermiticity of
the Hamiltonian).
For some reasons, the Lagrangians of this type and the associated quantum-mechanical systems never got any
detailed attention in the literature\footnote{The existence of such models was mentioned
in~\cite{CoPa,GiPaSt,Hull}.
In particular, it was noticed in~\cite{GiPaSt} that they cannot be obtained by dimensional reduction from
the Lorentz-covariant ${\cal N}=1$, $d=2$ sigma models, as distinct from those based on the
Lagrangians~\eqref{121stand} with symmetric metric (the ${\cal N}=2$, $d=1$ multiplets $({\bf 1},{\bf 2},{\bf 1})$ and
$({\bf 2},{\bf 2},{\bf 0})$ were termed in~\cite{GiPaSt} as $N=2a$ and $N=2b$, respectively).
The precise relation with the general ${\cal N}=1$, $d=2$ sigma model Lagrangian is given in Appendix~\ref{appendixA}.}.
Our study is intended to f\/ill up this gap.

There are relationships of our model and the SQM models previously considered.
In particular, we demonstrate that our model can be recovered by an appropriate Hamiltonian reduction from
the general complex ${\cal N}=2$ SQM model possessing some shift isometries and dealing with the manifolds
of the doubled real dimension.
So, from the geometric point of view, the considered class of ${\cal N}=2$ SQM models can be interpreted as
a~restriction of the Dolbeault complex to the real slice of the original complex target space.

Second, though the complex thus obtained does not coincide with the de Rham complex describing the standard
$({\bf 1},{\bf 2},{\bf 1})$ SQM models with symmetric target space metric, it can be obtained from the latter by
a~similarity transformation of the holomorphic supercharges.
We demonstrate it explicitly for the simplest 2-dimensional model of this type with the ``primordial''
metric $g_{MN} = \delta_{MN} + i\epsilon_{MN} b(x)$.
The supercharges that we derive are related by a~similarity transformation to the de Rham supercharges with
a~conformally f\/lat ``associated'' metric
\begin{gather}
h_{MN}=\sqrt{1-b^2}\delta_{MN}.
\label{metrconf}
\end{gather}
For this simple model, we derive the explicit expressions for the quantum supercharges and the Hamiltonian.
The latter {does not coincide} with the standard de Rham Hamiltonian (i.e.\
the covariant Laplacian acting on the forms).
This stems from the fact that the similarity transformations for $Q$ and $\bar Q$ that we use are
dif\/ferent (see equation~\eqref{simtran} below).
We demonstrate that the system exhibits hidden ${\cal N}=4$ supersymmetry at the quantum level: the
relevant energy levels reveal 4-fold degeneracy.
We dwell on a~special choice of the function $b(x)$ describing the dynamics on the 2-sphere~$S^2$.
The detailed analysis of this new ${\cal N}=2$ (and ${\cal N}=4$) SQM model is performed in the Appendix~\ref{appendixB}.
In a~certain limit, the Hamiltonian thus obtained is related by a~similarity transformation to the
square of the Dirac operator on $S^2$.

\section{The general model}\label{Se2}

Our starting point is the superf\/ield Lagrangian~\eqref{121stand} where the tensor $g_{MN}$ involves now
an antisymmetric part,
\begin{gather}\label{ishodnik}
g_{MN}=g_{(MN)}+i b_{[MN]},
\end{gather}
with real $g_{(MN)}$, $b_{[MN]}$.
Plugging there the superf\/ield~\eqref{XM}, we derive the component Lagrangian
\begin{gather}
L=\frac{1}{2}g_{(MN)}\big(\dot{x}^M\dot{x}^N+F^MF^N\big)+b_{[MN]}\dot{x}^M F^N+\frac i2g_{(MN)}
\big(\bar\psi^N\nabla{\psi}^M-\nabla\bar\psi^N{\psi}^M\big)
\nonumber
\\
\hphantom{L=}{}-\frac{1}{2}b_{[MN]}\big(\bar\psi^N\dot{\psi}^M-\dot{\bar\psi^N}{\psi}
^M\big)-\frac12\partial_P\partial_Q\big(g_{(MN)}+i b_{[MN]}\big)\psi^M\bar\psi^N\psi^P\bar\psi^Q
\nonumber
\\
\hphantom{L=}{}
+G_{M,PQ}F^M\psi^P\bar\psi^Q-\frac{1}{2}\left(\partial_M b_{[NP]}+\partial_N b_{[MP]}\right)\dot{x}
^P\psi^M\bar\psi{}^N,
\label{offL}
\end{gather}
where
\begin{gather}
G_{M,PQ}=\Gamma_{M,PQ}-\frac i2\left(\partial_M b_{[PQ]}+\partial_P b_{[QM]}+\partial_Q b_{[MP]}\right)\nonumber
\\
\phantom{G_{M,PQ}}{}
=\frac12\left(\partial_P g_{MQ}+\partial_Q g_{PM}-\partial_M g_{PQ}\right)
\label{Chris}
\end{gather}
are complex Christof\/fel symbols calculated for the complex metric~\eqref{ishodnik}, while $\Gamma_{M,
PQ}$ are the standard Christof\/fels for $g_{(MN)}$
\begin{gather*}
\Gamma_{M,PQ}=\frac12\big[\partial_P g_{(MQ)}+\partial_Q g_{(MP)}-\partial_M g_{(PQ)}\big],
\qquad
\nabla\psi^M=\dot{\psi}^M+\Gamma^M_{NQ}\dot{x}{}^N\psi^Q.
\end{gather*}
The Lagrangian~\eqref{offL} is invariant (modulo a~total derivative) under the following ${\cal N} = 2$
supersymmetry transformations
\begin{alignat}{3}
& \delta x^M=-\epsilon\psi^M+\bar\epsilon\bar\psi^M,
\qquad&&
\delta\psi^M=\bar\epsilon\big(i\dot{x}^M-F^M\big), &
\nonumber
\\
& \delta\bar\psi^M=-\epsilon\big(i\dot{x}^M+F^M\big),
\qquad &&
\delta F^M=i\big(\epsilon\dot{\psi}^M+\bar\epsilon\dot{\bar\psi}^M\big). &
\label{offTr}
\end{alignat}

It is worth noting that the whole Lagrangian~\eqref{offL} can be written through the Hermitian metric
$g_{MN}$ because it is the latter that enters the generalized superf\/ield Lagrangian~\eqref{121stand} we
started with.
However, the terms with the derivatives of $b_{[MN]}$ cannot be fully accommodated through the generalized
Christof\/fel symbols~\eqref{Chris}.
For instance, the full part bilinear in the fermionic f\/ields can be written as
\begin{gather*}
\frac i2g_{MN}\big(\bar{\psi}^N\dot{\psi}^M-\dot{\bar{\psi}}{}^N{\psi}
^M\big)-\frac i2 (\partial_S g_{PN}-\partial_N g_{SP} )\dot{x}{}^P\psi^S\bar{\psi}^N,
\end{gather*}
and the second piece {\it cannot} be written through $G_{M, PQ}$.
This means that the 2-form f\/ield $b_{[MN]}$ cannot be interpreted as a~potential of some closed torsion.
Respectively, the Lagrangian~\eqref{offL} and the original superf\/ield Lagrangian do not reveal any extra
target space gauge symmetry associated with $b_{[MN]}$.
The only target space gauge symmetry of~\eqref{121stand} with the Hermitian metric~\eqref{ishodnik} is the
dif\/feomorphisms
\begin{gather}
\delta X^M=\Lambda^M(X),
\qquad
\delta g_{MN}=-\partial_M\Lambda^Pg_{PN}-\partial_N\Lambda^P g_{MP},
\label{trgDiff}
\end{gather}
like in the case of symmetric metric.
It is curious that the Hermitian non-symmetric metric, with the antisymmetric part having no interpretation
as a~gauge f\/ield, was considered by Einstein and Strauss more than 60 years ago~\cite{Eins,Eins1} as the
basic ingredient of some generalization of the standard Einstein gravity (for a~recent renewal of interest
in such theories, see~\cite{Cham,Cham1} and references therein).

As one more comment, we mention the extreme possibility of the choice $g_{(MN)}=0$ in~\eqref{ishodnik}.
In this case, the superf\/ield Lagrangian is reduced to ${\sim}b_{[MN]}(X)D X^M \bar{D}X^N$, and the
component Lagrangian~\eqref{offL} with $g_{(MN)}=0$ provides a~new ${\cal N}=2$ supersymmetric extension of
the Chern--Simons quantum mechanics~\cite{ChSqm2,ChSqm,ChSqm1,HT}, based on the multiplet $({\bf 1},{\bf 2},{\bf 1})$
(in~\cite{HT} only the version built on the chiral $({\bf 2},{\bf 2},{\bf 0})$ multiplet was considered).
In such a~model, with non-degenerate~$b_{[MN]}$, the f\/ield $F^M$ plays the dynamical role of the
canonical momenta for $X^M$ and cannot be eliminated algebraically.
Though being worthy of detailed study, this option is beyond the scope of our consideration in the present
paper.
In what follows, we assume that $g_{(MN)}\neq 0$.

After elimination of the auxiliary f\/ield $F^M$ by its equation of motion
\begin{gather*}%\label{FM}
F^M=b^M_{P}\dot{x}^P-G^M_{PQ}\psi^P\bar\psi^Q
\end{gather*}
(with the indices being raised with $g^{(MP)}$), the Lagrangian acquires the form
\begin{gather}
L=\frac{1}{2}g_{(MN)}\dot{x}^M\dot{x}^N-\frac12g^{(MN)}\big(b_{[MP]}\dot{x}^P-G_{M,PQ}
\psi^P\bar\psi^Q\big)\big(b_{[NS]}\dot{x}^S-G_{N,ST}\psi^S\bar\psi^T\big)
\nonumber
\\
\hphantom{L=}{}
+\frac i2g_{(MN)}\big(\bar\psi^N\nabla{\psi}^M-\nabla\bar\psi^N{\psi}^M\big)-\frac12b_{[MN]}
\big(\bar\psi^N\dot{\psi}^M-\dot{\bar\psi^N}{\psi}^M\big)
\nonumber
\\
\hphantom{L=}{}
-\frac12\partial_P\partial_Q(g_{(MN)}+i b_{[MN]})\psi^M\bar\psi^N\psi^P\bar\psi^Q
-\frac{1}{2}(\partial_M b_{[NP]}+\partial_N b_{[MP]})\dot{x}^P\psi^M\bar\psi{}^N.\label{polnyjL}
\end{gather}
Note that the bosonic core of the Lagrangian~\eqref{polnyjL} is
\begin{gather}
L^{\rm bos}=\frac{1}{2}G_{(MN)}\dot{x}^M\dot{x}^N,
\qquad
G_{(MN)}=g_{(MN)}+b_{[MP]}g^{(PS)}b_{[SN]},
\label{Tarbos}
\end{gather}
i.e.\
the antisymmetric f\/ield $b_{[MP]}$ makes a~non-trivial contribution to the target space
metric\footnote{This metric resembles a~block of the generalized bosonic metric in the so called ``double
f\/ield theory'' (see, e.g.,~\cite{double}) and the metrics obtained via $T$-duality~\cite{Tdual}.
However, it dif\/fers from such metrics by the sign before the second term.
This important dif\/ference is related to the fact that our original non-symmetric metric~$g_{MN}$ is
Hermitian, while its counterparts in the references just mentioned are real, with the antisymmetric parts
being gauge f\/ields.}.

The Lagrangian~\eqref{polnyjL} is invariant, up to a~total derivative, under the nonlinear supersymmetry
transformations
\begin{gather*}%\label{onTr}
\delta x^M=-\epsilon\psi^M+\bar\epsilon\bar\psi^M,
\qquad
\delta\psi^M=\bar\epsilon\big(i\dot{x}^M-b^M_{P}\dot{x}^P+G^M_{PQ}\psi^P\bar\psi^Q\big),
\nonumber
\\
\delta\bar\psi^M=-\epsilon\big(i\dot{x}^M+b^M_{P}\dot{x}^P-G^M_{PQ}\psi^P\bar\psi^Q\big).
\end{gather*}
Using these transformations, one can derive the conserved N\"other supercharges
\begin{gather}
Q=\psi^M\left[\Pi_M-\frac i2\partial_M\left(g_{(NP)}+i b_{[NP]}\right)\psi^N\bar\psi^P\right],
\nonumber
\\
\bar Q=\bar\psi^M\left[\Pi_M+\frac i2\partial_M\left(g_{(NP)}-i b_{[NP]}\right)\psi^P\bar\psi^N\right],\label{Q}
\end{gather}
where
\begin{gather*}%\label{PiM}
\Pi_M=g_{(MN)}\dot{x}^N-g^{(PN)}b_{[PM]}\big(b_{[NS]}\dot{x}^S-G_{N,ST}\psi^S\bar\psi^T\big)
\\
\phantom{\Pi_M=}
{}-\frac{i}{2}\left(\partial_P g_{(QM)}-\partial_Q g_{(PM)}\right)\psi^P\bar\psi^Q-\frac{1}{2}
\left(\partial_P b_{[QM]}+\partial_Q b_{[PM]}\right)\psi^P\bar\psi^Q
\end{gather*}
is the canonical momentum of $x^M$, $\Pi_M = \frac{\partial L}{\partial \dot{x}^M}$.
It is worth noting that it is easier to derive the supercharges starting just from the of\/f-shell
Lagrangian~\eqref{offL} and transformations~\eqref{offTr}.
The f\/inal expressions for $Q$, $\bar Q$ do not involve the auxiliary f\/ield $F^M$ and have the same
form~\eqref{Q}.

When $b_{[MN]}$ is absent, the supercharges~\eqref{Q} acquire the form
\begin{gather}\label{Qbezb}
Q=\psi^M\left[\Pi_M+\frac i2\Gamma_{M,NP}\psi^N\bar\psi^P\right],
\qquad
\bar Q=\bar\psi^M\left[\Pi_M-\frac i2\Gamma_{M,NP}\psi^P\bar\psi^N\right].
\end{gather}
These are the supercharges of the standard de Rham ${\cal N}=2$ sigma model written in a~somewhat unusual
form\footnote{Surprisingly, we were not able to f\/ind such a~simple representation for the classical
supercharges of the $({\bf 1},{\bf 2},{\bf 1})$ ${\cal N}=2$ models in literature.}.
The {\it usual} form is
\begin{gather}\label{Qtangent}
Q=\psi^M\left[P_M-i\Omega_{M,AB}\psi_A\bar\psi_B\right],
\qquad
\bar Q=\bar\psi^M\left[P_M-i\Omega_{M,AB}\bar\psi_A\psi_B\right],
\end{gather}
where $A$, $B$ are the tangent space indices, $\psi_A = e_{AM} \psi^M$, $g_{MN}= e_{AM}e_{AN}$, and
\begin{gather}
\Omega_{M,AB}=e_{AN}\big(\partial_M e^N_B+\Gamma^N_{MT}e^T_B\big)
\label{Lorconn}
\end{gather}
are spin connections.

Note that the canonical momenta $P_M$ entering~\eqref{Qtangent} are obtained by the variation of the
Lagrangian over $\dot{x}_M$ with f\/ixed $\psi_A$, $\bar \psi_A$, while $\Pi_M$ appearing
in~\eqref{Q},~\eqref{Qbezb} is obtained by the variation with f\/ixed $\psi^M$, $\bar \psi^M$.
These two canonical momenta are related as~\cite{FIS},
\begin{gather}\label{PcherezPi}
P_M=\Pi_M-\frac{\partial\dot{\psi}_A}{\partial\dot{x}^M}\frac{\partial L}{\partial\dot{\psi}_A}
-\frac{\partial\dot{\bar\psi}_A}{\partial\dot{x}^M}\frac{\partial L}{\partial\dot{\bar\psi}_A}
=\Pi_M+\frac i2\left[\partial_M e_{AP}e_{AQ}-\partial_M e_{AQ}e_{AP}\right]\psi^P\bar\psi^Q.
\end{gather}

For the standard sigma model, the form~\eqref{Qtangent} is more convenient because it allows one to perform
the quantization rather straightforwardly.
Indeed, the ``f\/lat'' fermion variables $\psi_A$, $\bar\psi_A$ constitute, together with $x^M$, $P_M$, the
canonically conjugated pairs,
\begin{gather*}%\label{canonic}
\big\{\psi_A,\bar\psi_B\big\}_{\rm P.B.}=-i\delta_{AB},
\qquad
\big\{x^M,P_N\big\}_{\rm P.B.}=\delta_N^M,
\\
\big\{\psi_A,x^M\big\}_{\rm P.B.}=\big\{\bar\psi_A,x^M\big\}_{\rm P.B.}=\big\{\psi_A,P_M\big\}_{\rm P.B.}=\big\{\bar\psi_A,P_M\big\}_{\rm P.B.}=0.
\end{gather*}

When quantizing, we have to replace $P_M \to -i \partial/\partial_M$,
$\bar \psi_A \to \partial/\partial \psi_A$ and to choose a~parti\-cu\-lar way of ordering the momenta and
coordinates.
Especially convenient covariant nilpotent quantum supercharges have the {\it same} functional form
as~\eqref{Qtangent}\footnote{See~\cite{howto} for general recipes of resolving the ordering ambiguities in
SQM models.}.
They are Hermitian-conjugated to each other, bearing in mind the factor $\sqrt{g}$ in the measure,
\begin{gather}\label{conjug}
\bar Q=(\det g)^{-1/2}Q^\dagger(\det g)^{1/2},
\end{gather}
where $Q^\dagger$ is related to $Q$ by a~``naive'' conjugation.
Other orderings correspond to some extra conjugations, $Q \to e^W Q e^{-W}$, $\bar Q \to e^{-W} \bar Q e^W$.
This is interpreted as introducing potentials on the manifold~\cite{Witgeom,Witgeom1}.

The variables $\psi^M$, $\bar\psi^M$ satisfy more complicated relations,
\begin{gather*}%\label{skobki}
\begin{split}
& \big\{\psi^M,\bar\psi^N\big\}_{\rm P.B.}=-ig^{MN},
\\
& \big\{\Pi_M,\psi^N\big\}_{\rm P.B.}=-\frac12\partial_M g^{NQ}\psi_Q,
\qquad
\big\{\Pi_M,\bar\psi^N\big\}_{\rm P.B.}=-\frac12\partial_M g^{NQ}\bar\psi_Q.
\end{split}
\end{gather*}
In this case, the quantization procedure is somewhat trickier, with quantum commutators corresponding not
to the Poisson brackets, but rather to the Dirac brackets.

In our case with non-vanishing $b_{[MN]}$, we can also introduce canonically conjugated f\/lat fermion
variables related to $\psi^M$, $\bar\psi^M$ by {\it complex} vielbeins, and then quantize.
We will do this in the next section for the simplest nontrivial 2-dimensional model.

However, to derive the result that the model~\eqref{121stand} with generic Hermitian metric $g_{MN}$ can be
obtained by a~Hamiltonian reduction of a~certain Dolbeault sigma model (the subject of Section~\ref{Se4}), we do
not need to come to grips with quantization.
The whole reasoning will be carried out at the classical level and in this case the supercharges~\eqref{Q}
prove to be more convenient than the supercharges~\eqref{Qtangent}.

\section{Two-dimensional model. Similarity transformation}\label{Se3}

The simplest nontrivial case corresponds to just two bosonic coordinates,
$x^{M=1,2}$, with the f\/lat symmetric part of the metric, $g_{(MN)} = \delta_{MN}$.
In this case, all expressions are greatly simplif\/ied.
In particular, one can write $b_{[MN]} = b\epsilon_{MN}$.
The Christof\/fels~\eqref{Chris} vanish.
The quartic fermionic terms in the Lagrangian also vanish.
The latter acquires the form
\begin{gather}
L=\frac{1}{2}\big(1-b^2\big)\dot{x}^M\dot{x}^M-\frac{i}{2}\big(\delta_{MN}+ib\epsilon_{MN}
\big)\big(\dot{\psi}^M\bar\psi^N-{\psi}^M\dot{\bar\psi}^N\big)
\nonumber
\\
\phantom{L=}{}-\frac{1}{2}\left[\partial_M b\epsilon_{NP}+\partial_N b\epsilon_{MP}\right]\dot{x}
^P\psi^M\bar\psi{}^N.\label{Ldim2}
\end{gather}

We will assume $b^2 < 1$.
The degenerate case $b^2=1$ when the kinetic term vanishes is discussed at the end of this section.
The Lagrangian is then reduced to~\eqref{Lb1}.
When $b^2 >1$, the kinetic term has a~ghost signature, which requires a~special analysis.

The relevant classical supercharges read
\begin{gather}
\label{Q2mernoe}
Q=\psi^M\left[\Pi_M+\frac12\partial_M b\epsilon_{NP}\psi^N\bar\psi^P\right],
\qquad
\bar Q=\bar\psi^M\left[\Pi_M-\frac12\partial_M b\epsilon_{NP}\psi^N\bar\psi^P\right].
\end{gather}

In this case, the complex vielbeins transforming the variables $\psi^M$, $\bar\psi^M$ into canonically
conjugated pairs can be found explicitly,
\begin{gather}
\label{vielbeins}
\psi^M=e^M_A\psi_A,
\qquad
\bar\psi^M=\bar{e}^M_A\bar\psi_A,
\qquad
\psi_A=e_{AM}\psi^M,
\qquad
\bar\psi_A=\bar{e}_{AM}\bar\psi^M,
\\
e_{AM}=g_+\delta_{AM}-ig_-\epsilon_{AM},
\qquad
\bar e_{AM}=g_+\delta_{AM}+ig_-\epsilon_{AM},
\nonumber
\\
e_A^M=e_{AM}/\sqrt{1-b^2},
\qquad
\bar e_A^M=\bar e_{AM}/\sqrt{1-b^2},
\label{complviel}
\end{gather}
where
\begin{gather*}
g_\pm=\frac12\big(\sqrt{1+b}\pm\sqrt{1-b}\big).
\end{gather*}
These vielbeins satisfy the relations
\begin{gather*}%\label{vierb}
e_{AM}\bar{e}_{AN}=\delta_{MN}+ib\epsilon_{MN},
\qquad
e_{AM}e^M_B=\delta_{AB},
\qquad
e_{AM}e^N_A=\delta_{M}^N,
\qquad
\text{and \ \ c.c.}
\end{gather*}
The Lagrangian~\eqref{Ldim2} can then be rewritten as
\begin{gather}\label{Ldim2flat}
L=\frac{1}{2}\big(1-b^2\big)\dot{x}^M\dot{x}^M
-\frac{i}{2}\big(\dot{\psi}_A\bar\psi_A-{\psi}_A\dot{\bar\psi}_A\big)
-\frac{1}{2}[\partial_M b\epsilon_{NP}+\partial_N b\epsilon_{MP}]
\dot{x}^P e_A^M\bar e_B^N\psi_A\bar\psi_B
\end{gather}
(note that the vielbein time derivatives coming from the two parts of the fermion kinetic term are canceled
amongst each other).
With this form of the Lagrangian, it is obvious that $\psi_A$, $\bar\psi_B$ and $P_M$, $x^N$ indeed constitute
the mutually commuting canonical pairs.

The classical supercharges~\eqref{Q2mernoe} can be expressed via the fermion variables with the tangent
space indices.
The direct substitution gives
\begin{gather}
Q=\psi_A e^M_A\left[\Pi_M+\frac{\partial_M b}{2(1-b^2)}\left(\epsilon_{BC}+ib\delta_{BC}
\right)\psi_B\bar\psi_C\right],
\nonumber
\\
\bar Q=\bar\psi_A\bar e^M_A\left[\Pi_M-\frac{\partial_M b}{2(1-b^2)}\left(\epsilon_{BC}+ib\delta_{BC}
\right)\psi_B\bar\psi_C\right].\label{Q2mernoekas}
\end{gather}
An important point is that, in the considered case, due to the specif\/ic structure of the vielbeins, the
following equality holds
\begin{gather*}
P_M=\Pi_M+\frac i2\big[\big(\partial_M e^A_P\big)\bar e^A_Q-(\partial_M\bar e^A_Q)e^A_P\big]\psi^P\bar\psi^Q=\Pi_M.
\end{gather*}

It is a~straightforward exercise to make sure that these supercharges form the standard ${\cal N}=2$, $d=1$
superalgebra
\begin{gather*}%\label{N2class}
\{Q,Q\}_{\rm P.B.}=\{\bar Q,\bar Q\}_{\rm P.B.}=0,
\qquad
\{Q,\bar Q\}_{\rm P.B.}=-2i H^{\rm cl},
\end{gather*}
where $H^{\rm cl}$ is the classical Hamiltonian corresponding to the Lagrangian~\eqref{Ldim2}
(or~\eqref{Ldim2flat}),
\begin{gather}
H^{\rm cl}=\frac{1}{2(1-b^2)}\left[\Pi_M+\frac{1}{2}\left(\partial_Tb\epsilon_{NM}
+\partial_N b\epsilon_{TM}\right)\psi^T\bar\psi^N\right]^2.
\label{Hcl}
\end{gather}

We notice now that, using antisymmetry of the expressions like $\psi_A \psi_B$ under the permutation $A
\leftrightarrow B$ and 2-dimensional specif\/ics, one can get rid of the structure $\delta_{BC}$ in the
supercharges~\eqref{Q2mernoekas} and rewrite them as
\begin{gather}
Q=\psi_A e_A^M\left(\Pi_M-i\Omega_{M,BC}\psi_B\bar\psi_C\right),
\qquad
\bar Q=\bar\psi_A\bar{e}_A^M\left(\Pi_M-i\bar\Omega_{M,BC}\bar\psi_B\psi_C\right)
\label{Q3}
\end{gather}
with
\begin{gather}
\Omega_{M,BC}=\frac{\epsilon_{BC}}{2\big(1-b^2\big)}\left(b\epsilon_{NM}\partial_N b+i\partial_M b\right).
\label{spconn1}
\end{gather}

This object has an interesting geometric interpretation.
It is just the spin connection for the {\it complex} vielbeins $e_{AM}$, $e^N_B$ def\/ined
in~\eqref{complviel}.
It can be computed by the standard formulae
\begin{gather*}
\Omega_{M,AB}=e_{AN}\big(\partial_M e^N_B+\hat{\Gamma}^N_{MT}e^{T}_{B}\big)=e_{CM}\Omega_{C,AB},
\\
\Omega_{C,AB}=e^M_B e^N_A\partial_{[M}e_{CN]}+e^M_B e^N_C\partial_{[M}e_{AN]}
+e^M_C e^N_A\partial_{[M}e_{BN]}
\\
\hphantom{\Omega_{C,AB}}{}
=\frac{1}{2\big(1-b^2\big)}\epsilon_{AB}\left(ie^M_C\partial_M b+be^M_D\partial_M b\epsilon_{DC}\right).
\end{gather*}
Here, $\hat{\Gamma}^N_{MT}$ is the standard Levi-Civita connection for the real conformally f\/lat metric
$h_{MN} = e_{AM}e_{AN} = \sqrt{1-b^2}\delta_{MN}$ as in~\eqref{metrconf}.

Consider the {\it real part} of the spin connection~\eqref{spconn1}.
One observes that it can be interpreted as the {\it standard} spin connection~\eqref{Lorconn} for the same
conformally f\/lat metric $h_{MN}$, but with the naturally chosen {\it real} vielbeins
\begin{gather}
\label{vbconf}
\tilde e_{AM}=\big(1-b^2\big)^{1/4}\delta_{AM}.
\end{gather}
Being truncated in this way, the supercharges~\eqref{Q3} {\it coincide} with the supercharges of the
standard $({\bf 1},{\bf 2},{\bf 1})$ supersymmetric sigma model~\eqref{Qtangent}, though involving the new metric
$h_{MN}$, equation~\eqref{metrconf}, which is dif\/ferent from (but conformal to) the bosonic target metric
in~\eqref{Ldim2}.
So the quantum version of these supercharges describes the de Rham complex associated with the metric~$h_{MN}$.

Now, come back to the full expressions~\eqref{Q3}.
They are classical.
Once again, to obtain the quantum supercharges, one should order the fermion operators in a~certain
particular way.
Let us do it in the same way as for the standard de Rham supercharges.
We thus write
\begin{gather}
Q^{\rm qu}=-i\psi_A\big(e^M_A\partial_M+\Omega_{A,BC}\bar\psi_B\psi_C\big),
\qquad
\bar Q^{\rm qu}=-i\bar\psi_A\big(\bar{e}^M_A\partial_M+\bar\Omega_{A,BC}\psi_B\bar\psi_C\big).
\label{Qquan}
\end{gather}
One easily checks that these supercharges are nilpotent and hence satisfy the standard quantum ${\cal N}=2$
superalgebra.
The anticommutator $\{\bar Q^{\rm qu}, Q^{\rm qu}\}$ gives the quantum Hamiltonian $H^{\rm qu}$:
\begin{gather}
H^{\rm qu}=\frac{1}{2(1-b^2)}\bigg[{-}\partial_M\partial_M+\frac{1}{i} (\partial_Tb\epsilon_{NM}
+\partial_N b\epsilon_{TM} )e^T_A\bar e^N_B\psi_A\bar\psi_B\partial_M
\nonumber
\\
 \phantom{H^{\rm qu}=}{}+\frac{1}{2}
(\partial_Tb\partial_M b)e^T_A\bar e^M_B e^S_D\bar e^S_C\psi_A\psi_D\bar\psi_B\bar\psi_C\bigg]
\nonumber
\\
\phantom{H^{\rm qu}=}
{}-\frac{1}{2}\big(\bar{e}^M_A\partial_M e^N_A+e^N_A\bar\Omega_{C,CA}\big)\partial_N
+\frac{1}{2}\big(e^M_A\partial_M\bar\Omega_{C,BC}-\bar e^M_D\partial_M\Omega_{B,AD}
\big)\psi_A\bar\psi_B.
\label{Hq}
\end{gather}

An important observation is that the supercharges~\eqref{Qquan} can be obtained from the quantum
supercharges of the de~Rham complex with the metric~\eqref{metrconf} and vielbeins~\eqref{vbconf} by
a~similarity transformation
\begin{gather}\label{simtran}
Q^{\rm qu}=R^{-1}Q^{\text{de Rham}}R,
\qquad
\bar Q^{\rm qu}=R\bar Q^{\text{de Rham}}R^{-1},
\end{gather}
with
\begin{gather}\label{R}
R=\exp\left\{\frac i4\ln\frac{1+b}{1-b}\epsilon_{AB}\psi_A\bar\psi_B\right\}.
\end{gather}
The operator~\eqref{R} is not unitary, $R^\dagger = R$ rather than $R^{-1}$, and hence $Q^{\rm qu}$ and
$\bar Q^{\rm qu}$ are rotated in a~dif\/ferent manner.
The relations~\eqref{simtran},~\eqref{R} imply that the property~\eqref{conjug} that holds for the de Rham
supercharges holds as well for $Q^{\rm qu}$, $\bar Q^{\rm qu}$.
In other words, the supercharges~\eqref{simtran} are mutually Hermitian with the measure $\sqrt{h} =
\sqrt{1-b^2}$.

By construction, the Hamiltonian~\eqref{Hq} is Hermitian with the same measure.
It does not seem to be related, however, to the de~Rham Hamiltonian (the covariant Laplacian) by any
similarity transformation and has, as we will see, a~distinct spectrum.
In the sectors $F=0,2$, the Hamiltonian is reduced to the simple expressions
\begin{gather}
2H^{F=0}=\frac1{1-b^2}P_M^2-\frac{i b\partial_M b+\epsilon_{PM}\partial_P b}{(1-b^2)^2}P_M,
\nonumber
\\
2H^{F=2}=\frac1{1-b^2}P_M^2-\frac{i b\partial_M b-\epsilon_{PM}\partial_P b}{(1-b^2)^2}P_M.\label{HF=02}
\end{gather}
It is clear that for any eigenfunction $\Psi^{F=0}$ of $H^{F=0}$, the complex conjugate function
$\Psi^{F=2} = \big(\Psi^{F=0}\big)^\star$ is an eigenfunction of $H^{F=2}$ with the same eigenvalue.
Supersymmetry dictates that the states $Q^{\rm qu} \Psi^{F=0}$ and $\bar Q^{\rm qu} \Psi^{F=2}$ have also
the same energy.
Thus, in this case, the energy levels display a~4-fold degeneracy and, hence, the system enjoys in fact an
extended ${\cal N}=4$ supersymmetry.
One can remind here that any 2-dimensional manifold is K\"ahler and hence the de Rham complex can be
extended to the K\"ahler--de Rham complex with extended supersymmetry.
There is no reason to believe, however, that such an extended supersymmetry is also valid in the generic
higher-dimensional case.

Note that the ${\cal N}=2$ superf\/ield action corresponding to the Lagrangian~\eqref{121stand} with
$g_{MN} =\delta_{MN} + i b (X)\epsilon_{MN}$, $M,N=1,2$, reveals no any obvious second ${\cal N}=2$
supersymmetry which would complete the manifest ${\cal N}=2$ one to ${\cal N}=4$.
So the ${\cal N}=4$ supersymmetry we have observed either is realized by highly nonlinear superf\/ield
transformations involving the target potential $b(X)$, or is a~pure quantum-mechanical phenomenon, like
in~\cite{anomaly}.
In any case, the underlying ${\cal N}=4$ supermultiplet is an on-shell version
of the multiplet $({\bf 2},{\bf 4},{\bf 2})$, and its f\/ield content is the same
as for two ${\cal N}=2$ multiplets $({\bf 1},{\bf 2},{\bf 1})$\footnote{This ${\cal N}=2$ splitting
of the multiplet $({\bf 2},{\bf 4},{\bf 2})$, is dif\/ferent
from the splitting $({\bf 2},{\bf 4},{\bf 2}) = ({\bf 2},{\bf 2},{\bf 0}) \oplus ({\bf 0},{\bf 2},{\bf 2})$,
considered in~\cite{ISDir} and yielding the standard K\"ahlerian ${\cal N}=4$ supersymmetric mechanics.}.

A non-triviality of the quantum problem considered in this section
is the existence of {\it four} dif\/ferent metrics which should not be confused with each other.
For reader's convenience, we thus reiterate.
\begin{itemize}\itemsep=0pt
\item First, there is the primordial complex Hermitian metric~\eqref{ishodnik}.
Its geometric meaning will be clarif\/ied in the next section.
\item The symmetric part $g_{(MN)}$ of this complex metric does not have a~special meaning (at least, we do
not see it), but it enters the formulae~\eqref{polnyjL},~\eqref{Tarbos} in Section~\ref{Se2}.
\item Next, there is the ``kinetic metric''~\eqref{Tarbos} that determines the bosonic dynamics.
It is constructed from both $g_{(MN)}$ and the antisymmetric potential $b_{[MN]}$.
In the two-dimensional case, it boils down to $ h_{MN}^{\rm kinetic} = (1-b^2) \delta_{MN}$.
\item Finally, there is an ``associated metric''~-- the metric of the de Rham complex with supercharges
related to our supercharges by a~similarity transformation.
In the 2-dimensional case, we have derived $ h_{MN}^{\rm associated} = \sqrt{1-b^2}  \delta_{MN}$.
\end{itemize}

A~detailed analysis of the spectrum of the Hamiltonian~\eqref{Hq} in the simple case when the kinetic
metric $h_{MN} \propto \delta_{MN} /(1+x_P^2)^2$ describes the 2-sphere $S^2$ is performed in the Appendix~\ref{appendixB}.

As was mentioned above, the whole consideration above was performed assuming $b^2 < 1$.
When $b^2 >1$, the spectrum of the Hamiltonian does not have a~bottom (but is probably bounded from above).
At the points $b = \pm 1$, the bosonic metric in~\eqref{Ldim2} vanishes and, furthermore, the matrices
$\delta_{MN} \pm i\epsilon_{MN}$ entering the fermionic terms get degenerate.
It is interesting to see what happens with the original superf\/ield Lagrangian at these special points.
Without loss of generality, we can choose, e.g., $b= 1$.
The Lagrangian~\eqref{121stand} with $g_{MN} = \delta_{MN} + i\epsilon_{MN}$, $M,N = 1,2$, is reduced to
\begin{gather}
L_{(b=1)}=-\frac12\int d\theta d\bar\theta\, DZ\bar D\bar Z,
\qquad
Z:=X^1-iX^2,
\qquad
\bar Z=X^1+iX^2.
\label{b1}
\end{gather}
The (anti)chiral spinor superf\/ields $\Psi:= DZ$, $\bar\Psi = -\bar D\bar Z$,  $D\Psi = \bar D \bar\Psi =
0$, are just the superf\/ield strengths of the purely fermionic ${\cal N}=2$ supermultiplet (${\bf 0},{\bf 2},
{\bf 2}$), with $Z$, $\bar Z$ being the relevant gauge prepotentials exhibiting the gauge freedom
\begin{gather*}
Z\rightarrow Z+D\Omega,
\qquad
\bar Z\rightarrow\bar Z-\bar D\bar\Omega,
\qquad
\Omega=\omega+\theta\phi+\bar\theta\sigma+\theta\bar\theta\delta.
\end{gather*}
Here, $\omega$, $\delta$ and $\phi$, $\sigma$ are arbitrary fermionic and bosonic complex
functions\footnote{The superf\/ield gauge parameter $\Omega$ itself is def\/ined up to addition of an
arbitrary antichiral superf\/ield, so only the gauge parameters $\phi$ and $\delta$ actually matter.}.
Using this gauge freedom, one can choose the ``Wess--Zumino gauge''
\begin{gather*}
Z=\theta\lambda+\theta\bar\theta g,
\qquad
\bar Z=-\bar\theta\bar\lambda+\theta\bar\theta\bar g,
\end{gather*}
in which the Lagrangian~\eqref{b1} is reduced to
\begin{gather}\label{Lb1}
L_{(b=1)}=\frac i2\big(\bar\lambda\dot\lambda-\dot{\bar\lambda}\lambda\big)+\frac12g\bar g,
\end{gather}
that is the free Lagrangian of the multiplet (${\bf 0},{\bf 2},{\bf 2}$).
Thus at the points $b =\pm1$ we end up with the (${\bf 0},{\bf 2},{\bf 2}$) Lagrangian.
The original two (${\bf 1},{\bf 2},{\bf 1}$) multiplets are combined into a~complex gauge bosonic superf\/ield which
provides an alternative description of the (${\bf 0},{\bf 2},{\bf 2}$) multiplet.
On shell $g =0$, and the relevant Hamiltonian and supercharges vanish.

\section{Hamiltonian reduction}\label{Se4}

The geometric meaning of the model considered can be further clarif\/ied by addressing a~special class of
{\it complex} $n$-dimensional manifolds with the Hermitian metric
\begin{gather*}%\label{complmetr}
h_{j\bar k}=g_{(jk)}+ib_{[jk]},
\qquad
(h_{j\bar k})^\dagger=h_{k\bar j},
\end{gather*}
and constraining the real functions $ g_{(j k)}$, $b_{[j k]}$ to depend {\it only} on the real parts of the
complex variables $z^j$ $(j=1,\ldots, n)$.

As was mentioned above, one can def\/ine on complex manifolds ${\cal N}=2$ supersymmetric sigma
models~\cite{Hull,ISDir} whose supercharges are isomorphic to the exterior holomorphic derivative and its
conjugate and realize thereby Dolbeault complex.
The underlying ${\cal N}=2$ multiplets are of the (${\bf 2}, {\bf 2}, {\bf 0}$) type, such that each
complex bosonic coordinate $z^j$ has one complex fermionic superpartner $\psi^j$.
In other words, Dolbeault sigma model for a~manifold of complex dimension $n$ has the same number of
holomorphic fermion variables as de Rham sigma model or quasicomplex sigma model on a~manifold of real
dimension~$n$.
Hilbert spaces in the former and in the latter are therefore tightly connected.

For sure, the mapping between the $({\bf 2},{\bf 2},{\bf 0})$ models and the Dolbeault complexes known to
mathematicians and similarity between Hilbert spaces of dif\/ferent complexes concern {\it quantum}
supercharges in the f\/irst place.
However, in order to establish the correspondence between our supercharges and the Dolbeault supercharges,
it is suf\/f\/icient to compare only the {\it classical} versions of the two sets.

We introduce holomorphic vielbeins satisfying\footnote{We follow the notation of Section~2
of~\cite{ISDir}, where the reader is redirected for further details.}
\begin{gather*}
e_k^a e^{\bar a}_{\bar i}=h_{k\bar i},
\qquad
e^k_a e^{\bar i}_{\bar a}=h^{\bar i k},
\qquad
h^{\bar i k}h_{k\bar j}=\delta^{\bar i}_{\bar j},
\qquad
h_{k\bar j}h^{\bar j l}=\delta^l_k,
%\label{viel}
\\
e_k^ae_a^j=\delta^j_k,
\qquad
e^k_ae^b_k=\delta^b_a,
\qquad
e_{\bar k}^{\bar a}e_{\bar a}^{\bar j}=\delta^{\bar j}_{\bar k},
\qquad
e^{\bar k}_{\bar a}e^{\bar b}_{\bar k}=\delta^{\bar b}_{\bar a}.
%\label{ort}
\end{gather*}
The Dolbeault ${\cal N}=2$ supercharges were written in~\cite{ISDir} in the form analogous
to~\eqref{Qtangent}\footnote{See equation~(3.15) of~\cite{ISDir}, where one has to set $W=0$.},
\begin{gather}\label{Qcompltangent}
Q=\psi^k\big(P_k-i\bar\psi^{\bar a}\psi^b\Omega_{k,{\bar a}b}\big),
\qquad
\bar Q=\bar\psi^{\bar k}\big(P_{\bar k}+i\bar\psi^{\bar a}\psi^b\bar\Omega_{{\bar k},b{\bar a}}\big),
\end{gather}
where $\Omega_{j, {\bar a} b} = e^a_p (\partial_j e^p_b + \Gamma^p_{jk} e^k_b )$ are complex spin
connections and the momentum $P_M$ is the canonical momentum calculated at f\/ixed $\psi^a$, $\bar\psi^{\bar
a}$.
For further convenience, we omitted the factors $\sqrt{2}$ compared to the expressions in~\cite{ISDir}, so
the relevant Hamiltonians coincide with each other up to the factor~$1/2$.

To establish the sought correspondence with the quasicomplex model of the preceding sections, it is
convenient to rewrite the supercharges~\eqref{Qcompltangent} in terms of the fermionic variables with the
world indices, like in~\eqref{Q},~\eqref{Qbezb}.
We obtain
\begin{gather}\label{Qcomplworld}
Q=\psi^j\left(\Pi_j+\frac i2\Gamma_{j,k{\bar p}}\psi^k\bar\psi^{\bar p}\right),
\qquad
\bar Q=\bar\psi^j\left(\Pi_{\bar j}-\frac i2\Gamma_{{\bar j},k{\bar p}}\psi^k\bar\psi^{\bar p}\right).
\end{gather}
Here,
\begin{gather}
\Gamma_{j,k{\bar p}}=\frac{1}{2}(\partial_k h_{j{\bar p}}-\partial_j h_{k{\bar p}}),
\qquad
\Gamma_{{\bar j},k{\bar p}}=\frac12(\partial_{\bar p}h_{k{\bar j}}-\partial_{\bar j}h_{k{\bar p}}),
\label{ChristofN2}
\end{gather}
and the momenta $\Pi_j$, $\Pi_{\bar j}$ are related to $P_j$, $P_{\bar j}$ in the same way as
in~\eqref{PcherezPi},
\begin{gather}\label{PcherezPicompl}
P_j=\Pi_j + \frac i2\big[(\partial_j e^a_p)\bar e^{\bar a}_{\bar q} - (\partial_j e^{\bar a}_{\bar q}
)e^a_p\big]\psi^p\bar\psi^{\bar q},
\qquad
P_{\bar j}=\Pi_{\bar j} + \frac i2\big[(\partial_{\bar j}e^a_p)\bar e^{\bar a}_{\bar q} - (\partial_{\bar j}
e^{\bar a}_{\bar q})e^a_p\big]\psi^p\bar\psi^{\bar q}.\!\!\!
\end{gather}
Note that the Christof\/fel symbols~\eqref{ChristofN2} {\it vanish} for K\"ahler manifolds, killing the
three-fermion terms in~\eqref{Qcomplworld}.
This nice property is specif\/ic for the form~\eqref{Qcomplworld} of the supercharges.

Requiring the metric and vielbeins to depend only on the real parts of~$z^j$ amounts to assuming that the
manifold involves isometries realized as shifts in the imaginary directions.
One can observe that, in this case, the Poisson brackets of the supercharges (and hence of the Hamiltonian)
with the generator of these imaginary shifts $\Pi_j - \Pi_{\bar j} = P_j - P_{\bar j}$ all vanish.
This allows one to perform a~Hamiltonian reduction~-- to identify $\Pi_j \equiv \Pi_{\bar j}$ and to
forget about the imaginary parts of the coordinates whatsoever.

Physically, the picture becomes more transparent at the quantum level~-- one can observe that the metric
isometries allow one to def\/ine the Hamiltonian acting on the restricted Hilbert space, with the wave
functions depending only on ${\rm Re}\,z^j$.\footnote{In fact, the conditions $\Pi_j - \Pi_{\bar j} = 0$
represent f\/irst class constraints, like the Gauss law constraint in standard gauge theories.
Our reduced system~\eqref{polnyjL} can thus be interpreted as a~certain gauge model with the constraints
resolved.
At the level of of\/f-shell superf\/ield actions, the same reduction could presumably be accomplished using
the gauging techniques of~\cite{DeldIv1}.}
But, for establishing the correspondence, one can stay at the classical level.

One has a~pleasure to observe that, after this identif\/ication, the expressions~\eqref{Qcomplworld}
coincide with~\eqref{Q}.
In particular, the 2-dimensional model of the previous section can be obtained by the Hamiltonian reduction
from the complex Dolbeault model living on a~manifold of complex dimension~2 with the metric $h_{j\bar k} =
\delta_{j k} + i b \epsilon_{j k}$ and the vielbeins~\eqref{vielbeins}.

For completeness, we will f\/inally discuss the same reduction directly in terms of Hamiltonians.
The classical Hamiltonian of the complex ${\cal N}=2$ model (with the vanishing background gauge potential,
i.e.\
with $W =0$) was written in~\cite{ISDir} as
\begin{gather*}
H_{\rm cl}=h^{\bar{k}j}\big(P_j+i\hat{\Omega}_{j,\bar b a}\psi^a\bar\psi^{\bar b}\big)\big(P_{\bar k}
-i\hat{\bar\Omega}_{\bar k,c\bar d}\psi^c\bar\psi^{\bar d}\big)-e^t_a e^j_c e^{\bar l}_{\bar b}e^{\bar k}
_{\bar d}(\partial_t\partial_{\bar l}h_{j\bar k})\psi^a\psi^c\bar\psi^{\bar b}\bar\psi^{\bar d},
\end{gather*}
where
\begin{gather*}
\hat{\Omega}_{j,\bar b a}=-\hat{\Omega}_{j,a\bar b}=e^b_k\partial_j e^k_a+e^{\bar t}_{\bar b}
e^k_a\hat{\Gamma}_{\bar t,jk},
\qquad
\hat{\Gamma}_{\bar t,jk}=\partial_k h_{j\bar t}.
\end{gather*}
After passing everywhere to the world indices (in particular, by making use of the
relation~\eqref{PcherezPicompl}) and identifying $\partial_j = \partial_{\bar j}$, this Hamiltonian can be
rewritten as
\begin{gather}
H_{\rm cl}=h^{\bar{k}j}
\left[\Pi_j+i\left(\partial_p h_{j\bar q}-\frac12\partial_j h_{p\bar q}\right)\psi^p\bar\psi^{q}\right]
\left[\Pi_k-i\left(\partial_p h_{q\bar k}-\frac12\partial_k h_{q\bar p}\right)\psi^q\bar\psi^{p}\right]
\nonumber
\\
\phantom{H_{\rm cl}=}
{}-\partial_t\partial_{l}h_{j\bar k}\psi^t\psi^j\bar\psi^{l}\bar\psi^{k}.
\label{Hamred}
\end{gather}
For the $2$-dimensional target metric $h_{j\bar k} = \delta_{j k} + i b \epsilon_{j k}$ the last term
in~\eqref{Hamred} vanishes and $H_{\rm cl}$ nicely coincides with~\eqref{Hcl}, up to the overall factor
$1/2$.

\section{Summary and outlook}

In this paper, we introduced and studied a~new class of ${\cal N}=2$
supersymmetric quantum mechanical systems, the quasicomplex quantum mechanics.
Its superf\/ield Lagrangian involves, besides the standard metric term, also an antisymmetric tensor which
{\it cannot} be identif\/ied with any torsion potential.
These two terms are naturally joined into a~non-symmetric Hermitian target space superf\/ield metric.
In components, the antisymmetric tensor generates some non-trivial target bosonic metric even in the case
when the standard metric is f\/lat.

From the geometrical point of view, these models realize a~new complex which coincides neither with the de
Rham complex nor with the Dolbeault one and seems not to be discussed earlier by mathematicians.
However, they are still related in a~certain way to both these complexes.
\begin{itemize}
\itemsep=0pt \item First, as we have seen, a~quasicomplex sigma model living on a~real $n$-dimensional
manifold can be reproduced through Hamiltonian reduction from the Dolbeault ${\cal N}=2$ models living on
a~manifold of complex dimension $n$ and possessing appropriate isometries.
The latter allow one to get rid of a~half of real bosonic coordinates.
\item Second, they may be related to a~certain de Rham complex through a~similarity transformation of the
{\it holomorphic} quantum supercharges.
\end{itemize}

We explicitly constructed here such a~transformation in the simplest two-dimensional examp\-le.
However, as was recently shown~\cite{taming}, {\it both} the generic multidimensional quasicomplex
system~\eqref{Q} {\it and} the standard de Rham system~\eqref{Qbezb} can be related to a~{\it free} system
by the proper similarity transformations of the supercharges.
Thus, a~combination of these transformations gives a~generic similarity transformation {\it quasicomplex
$\to$ de~Rham}, a~multidimensional ge\-ne\-ralization of the transformation~\eqref{simtran}.
Perhaps it is worth mentioning once more that this {\it similarity} of the two systems does not imply their
{\it equivalency}: their Hamiltonians do not coincide and have dif\/ferent spectra.

We studied in detail the simplest 2-dimensional version of these unusual models and found that, at the
quantum level, the spectrum involves a~4-fold degeneracy of the states, thus exhibiting a~hidden ${\cal
N}=4$ supersymmetry.
Studying (in Appendix~\ref{appendixB}) the spectrum further, we discovered its rather interesting features.
In a~certain limit, the spectrum consists of $\mathfrak{su}(2)$ multiplets with half-integer momenta in both fermionic
and {\it bosonic} sectors.
The Hamiltonian is related by a~similarity transformation to the square of the Dirac operator, $H =
/\!\!\!\!{\cal D}^2$.\footnote{We want to stress that in {\it this} case we are talking about the similarity
transformation of the {\it Hamiltonians} that gives the equivalence of the spectra, mapping of the wave
functions, etc.}

Studying the spectra of more complicated quasicomplex Hamiltonians in higher dimensions would be highly
desirable.

\appendix

\section[Relation to ${\cal N}=1$, $d=2$ sigma model]{Relation to $\boldsymbol{{\cal N}=1}$, $\boldsymbol{d=2}$ sigma model}\label{appendixA}

Here we will consider the one-dimensional reduction of the most general superf\/ield Lagrangian of the
${\cal N}=1$, $d=2$ sigma model associated with the real superf\/ield $X^M$.

Following~\cite{GiPaSt}, such a~Lagrangian in the light-cone parametrization of the ${\cal N}=1$, $d=2$
superspace can be written as
\begin{gather}
L_{(d=2)}=\int d\theta^+d\theta^-\Big\{G_{(MN)}D_+X^M D_-X^N+B_{[MN]}D_+X^M D_-X^N
\nonumber
\\
\hphantom{L_{(d=2)}=}{}
+G^{--}_{[MN]}D_-X^M D_-X^N+G^{++}_{[MN]}D_+X^M D_+X^N\Big\},\label{d2Act}
\end{gather}
where $\theta^\pm$ are real Grassmann coordinates, $D_{\pm} = \frac{\partial}{\partial \theta^{\pm}} +
i\theta^\pm\partial_{\pm\pm}$, $(D_\pm)^2 = i\partial_{\pm\pm}$, $\overline{(D_\pm)} = - D_\pm$, and
$G_{(MN)}(X)$, $B_{[MN]}(X)$, $G^{\pm\pm}_{[MN]}(X)$ are real functions.
Since $X^M$ are $d=2$ Lorentz scalars, the last two terms in~\eqref{d2Act} explicitly break $d=2$ Lorentz
covariance.
The Lagrangian~\eqref{d2Act} exhibits invariance under the target space dif\/feomorphisms realized like
in~\eqref{trgDiff} and gauge transformations associated with the antisymmetric f\/ield~$B_{[MN]}$
\begin{gather}
\delta B_{[MN]}(X)=\partial_M A_N(X)-\partial_N A_M(X),
\label{gaugeBMN}
\end{gather}
where $A_N(X)$ are arbitrary real parameters.
The $d=2$ Lorentz breaking tensors $G^{\pm\pm}_{[MN]}$ do not introduce any additional target space gauge
freedom.

Now we pass to the complex coordinates $\theta = \theta^+ + i\theta^-$, $\bar\theta = \theta^+ - i\theta^-$,
and the complex spinor derivatives
\begin{gather}
D_+=D-\bar D,
\qquad
D_-=i(D+\bar D),
\nonumber\\
\{D,D\}=\{\bar D,\bar D\}=\frac{i}{2}(\partial_{++}-\partial_{--}),
\qquad
\{D,\bar D\}=-\frac{i}{2}
(\partial_{++}+\partial_{--}).
\label{d2alg}
\end{gather}
In terms of $D$, $\bar D$ the Lagrangian~\eqref{d2Act}, up to an overall numerical coef\/f\/icient, can be
rewritten~as
\begin{gather}
\int d\bar\theta d\theta\bigg\{ (g_{(MN)}+i b_{[MN]} )D X^M\bar D X^N
\nonumber
\\
 \qquad{}
+\frac12{\cal B}_{[MN]}D X^M D X^N-\frac12\bar{{\cal B}}_{[MN]}\bar D X^M\bar D X^N\bigg\},
\label{d2Act1}
\end{gather}
where
\begin{gather*}
g_{(MN)}=G_{(MN)},
\qquad
b_{[MN]}=G^{++}_{[MN]}+G^{--}_{[MN]},
\\
{\cal B}_{[MN]}=B_{[MN]}-i\big(G^{++}_{[MN]}-(G^{--}_{[MN]}\big),
\qquad
\bar{{\cal B}}_{[MN]}=B_{[MN]}+i\big(G^{++}_{[MN]}-(G^{--}_{[MN]}\big).
\end{gather*}
After performing the $d=2 \rightarrow d=1$ dimensional reduction as $\partial_{++} = \partial_{--} =
-2\partial_t$, the spinor derivatives~\eqref{d2alg} become the ${\cal N}=2$, $d=1$ spinor derivatives
and~\eqref{d2Act1} is recognized as the Lagrangian~\eqref{121stand} with the non-symmetric Hermitian
metric~\eqref{ishodnik} modif\/ied by terms with extra antisymmetric tensor f\/ields ${\cal B}_{[MN]}$,
$\bar{{\cal B}}_{[MN]}$.
These objects are gauge potentials of the torsion on the target space~\cite{FIS}, with the same gauge
transformation law as in~\eqref{gaugeBMN}.
The f\/ield $b_{[MN]}$ does not bring in any new target space gauge freedom like its $d=2$ prototypes
$G^{\pm\pm}_{[MN]}$.

\section[Hamiltonian~(\ref{Hq}) on $S^2$ and its spectrum]{Hamiltonian~(\ref{Hq}) on $\boldsymbol{S^2}$ and its spectrum}\label{appendixB}

Consider f\/irst the Hamiltonian in the sector $F=0$ written in equation~\eqref{HF=02}.
Introduce the complex coordinates,
\begin{gather*}
\partial_1=\frac{\mu}{\sqrt{2}}\left(\partial_w+\partial_{\bar w}\right),
\qquad
\partial_2=\frac{i\mu}{\sqrt{2}}\left(\partial_{\bar w}-\partial_{w}\right).
\end{gather*}
The $S^2$ case is obtained by identifying
\begin{gather*}
1-b^2=\frac{\rho}{(1+w\bar w)^2}
\end{gather*}
($\mu$ and $\rho$ are arbitrary real constants).
This implies
\begin{gather}
b=\pm\frac{\sqrt{(1+X)^2-\rho}}{1+X},
\qquad
b'=\pm\frac{\rho}{(1+X)^2\sqrt{(1+X)^2-\rho}},
\qquad
X:=w\bar w.
\label{solb}
\end{gather}
Choosing in~\eqref{solb} the positive sign and substituting it into~\eqref{HF=02}, we obtain the
Hamiltonian
\begin{gather}
H_{S^2}=\frac{\mu^2}{\rho}\left[-(1+w\bar w)^2\partial\bar\partial-\frac{1}{2}
(1+w\bar w)(w\partial+\bar w\bar\partial)+\frac{1}{2}\frac{(1+w\bar w)^2}{\sqrt{(1+w\bar w)^2-\rho}}
J\right],
\label{S2ham}
\end{gather}
where
\begin{gather}\label{charge}
J=w\partial-\bar w\bar\partial
\end{gather}
is the charge operator which commutes with the whole $H_{S^2}$.

The Hamiltonian~\eqref{S2ham} is Hermitian with the measure $\sqrt{1-b^2} \propto 1/(1+ w\bar w)$ (see the
remark after~\eqref{R}).
Thus, the Hilbert space where this Hamiltonian acts involves the wave functions normalized as
\begin{gather}\label{norm}
\int |\Psi(w,\bar w) |^2\frac{dw d\bar w}{1+w\bar w}=1.
\end{gather}
It is $1+ w\bar w $ downstairs, not $(1 + w\bar w )^2$ as for the standard Laplacian on $S^2$!

In the sector $J=0$, the Hamiltonian acquires a~particularly simple form, such that the spectrum and wave
functions can be found analytically.
Indeed, introducing the variable $z = \frac {1-X}{1+X}$ (it is none other than the cosine of the polar
angle on $S^2$), the Schr\"odinger equation acquires in this sector the form
\begin{gather}
\big(z^2-1\big)\Psi^{\prime\prime}(z)+(z+1)\Psi^{\prime}(z)=E\Psi(z).
\label{ep01}
\end{gather}
The solutions are Jacobi polynomials,
\begin{gather*}
\Psi_m(z)=(1+z)P_m^{0,1}(z),
\qquad
m=1,\ldots,
\end{gather*}
which gives the spectrum
\begin{gather}
E_m=m^2.
\label{specJ0}
\end{gather}
Note that we have {\it excluded} the function $\Psi = {\rm const}$ (which is a~formal solution of~\eqref{ep01}),
because it is not normalizable with the measure in~\eqref{norm}.
This means that the zero-energy states are absent in the spectrum and supersymmetry is thus broken.
The f\/irst few states are
\begin{gather}%\label{firstpsi}
\Psi_1(z)=1+z,
\qquad
\Psi_2(z)=(1+z)(1-3z),
\nonumber\\
\Psi_3(z)=(1+z)\big(1+2z-5z^2\big),
\qquad
\ldots.
\label{psi0}
\end{gather}

When $J\neq 0$, the situation is more complicated.
The relevant wave functions are eigenfunctions of $J$ with integer non-zero eigenvalues.
We pose
\begin{gather*}
\Psi_J=w^J f_J(z)(J>0),
\qquad
\Psi_J=\bar w^{|J|}g_J(J<0).
\end{gather*}

The Schr\"odinger equation for $\Psi_{J>0}$ is reduced to the following equation for $f_J$,
\begin{gather}
\big(z^2-1\big)f_J^{\prime\prime}+(1+2J+z)f_J^{\prime}-\frac{J}{1+z}\left(1-\frac{1}
{\sqrt{1-\rho\frac{(1+z)^2}{4}}}\right)f_J=Ef_J.
\label{ep1}
\end{gather}
The function $g_J$ satisf\/ies the similar equation with $J$ being replaced by $|J|$ and with the opposite
sign before the second term in the round brackets.
These equations can be solved numerically, when imposing proper boundary conditions.

In the sector $F=2$, the Hamiltonian is the same as in~\eqref{S2ham} up to the opposite sign for the last
term.
The spectrum is the same as for $F=0$, but with the complex conjugated wave functions.
The states in the sector $F=1$ are obtained by the action of the supercharges.

Notice now that the problem is drastically simplif\/ied in the limit $\rho, \mu \to 0$, the ratio
$\mu^2/\rho$ being kept f\/ixed.
Let the latter be 1.
The Hamiltonian acquires the form
\begin{gather}
H_{(\rho=0)}=-(1+w\bar w)^2\partial\bar\partial-(1+w\bar w)\bar w\bar\partial.
\label{rho0}
\end{gather}
One can observe now that this Hamiltonian commutes not only with~\eqref{charge}, but also with the
operators
\begin{gather}\label{Jpm}
J_+=\partial+\bar w^2\bar\partial,
\qquad
J_-=\bar\partial+w^2\partial+w.
\end{gather}
They are mutually (anti)conjugated with respect to the measure ${\sim}\frac{1}{1 + w\bar w}$ and form the
$\mathfrak{su}(2)$ algebra,
\begin{gather*}
[J_+,J_-]=2J_3,
\qquad
[J_\pm,J_3]=\pm J_\pm,
\end{gather*}
with $J_3 = J + 1/2$.

This means that the eigenstates of~\eqref{rho0} represent ${\rm SU}(2)$ multiplets.
An amusing fact is that, in contrast to the case of the ordinary Laplacian, $\triangle = (1+ \bar w w)^2
\partial \bar \partial$, these multiplets correspond to half-integer momenta.
Indeed, in the ground state (with $E=1$), we f\/ind a~{\it doublet},
\begin{gather}%\label{2plet}
\Psi_0=\frac1{1+w\bar w},
\qquad
\Psi_+=\frac{\bar w}{1+w\bar w}.
\label{12}
\end{gather}
These states are obtained from one another by the action of $J_+$ and $J_-$.
At the level $E=4$, we have four states,
\begin{gather}%\label{4plet}
\Psi_-=\frac w{(1+w\bar w)^2},
\qquad\!\!\!\!
\Psi_0=\frac{1-2w\bar w}{(1+w\bar w)^2},
\qquad\!\!\!\!
\Psi_+=\frac{\bar w(2-w\bar w)}{(1+w\bar w)^2},
\qquad\!\!\!\!
\Psi_{++}=\frac{\bar w^2}{(1+w\bar w)^2}.\!\!\!\!\!
\label{32}
\end{gather}

Note that the $J=0$ reduction of the general Hamiltonian~\eqref{S2ham} exactly coincides with the reduction
of~\eqref{rho0} since $w\partial_w = \bar w \partial_{\bar w}$ on the $J=0$ wave functions.
Hence~\eqref{rho0} has the same energy spectrum given by~\eqref{specJ0} and all neutral components of the
above ${\rm SU}(2)$ multiplets are simultaneously eigenfunctions of the generic $J=0$ Hamiltonian.
In particular, the neutral functions $\Psi_0$ in~\eqref{12} and~\eqref{32} coincide, up to numerical
coef\/f\/icients, with $\Psi_1$ and $\Psi_2$ in the sequence~\eqref{psi0}.
Posing $m = s+1/2$, we represent the spectrum~\eqref{specJ0} as
\begin{gather}\label{specs}
E_s=s(s+1)+1/4
\end{gather}
with the multiplicity $2s+1$.
The Hamiltonian~\eqref{rho0} represents the Casimir operator of the algebra $\{J_\pm, J_3\}$ shifted by
a~constant.
Indeed, it is easy to directly check that for the realization~\eqref{Jpm}
\begin{gather*}
C_2=-\frac{1}{2}\big[{J}_+{J}_{-}+{J}_{-}{J}_{+}-2({J}_3)^2\big]=H_{(\rho=0)}-\frac{1}{4}.
\end{gather*}

One can recall now that the spectrum~\eqref{specs} is also characteristic for the Dirac operator on $S^2$
in the case when the gauge f\/ield is absent and only spin connections are taken into account (see,
e.g.,~\cite{Persik}).
And, indeed, one can observe that the Hamiltonian~\eqref{rho0} is related to the square of the Dirac
operator multiplied by the positive chirality projector, $H_D^{(+)} = /\!\!\!\!{\cal D}^2 \frac {1+\sigma_3}2$
and expressed via complex variables $w$, $\bar w $ by a~similarity transformation,
\begin{gather*}
H_{(\rho=0)}=\frac1{\sqrt{1+w\bar w}}H_D^{(+)}\sqrt{1+w\bar w}.
\end{gather*}
This transformation changes the measure: the eigenfunctions of $H_D$ are normalized with the standard
measure $\propto 1/(1+ w \bar w)^2$, while the eigenfunctions of $H_{(\rho = 0)}$ are normalized as
in~\eqref{norm}.
For instance, the wave functions~\eqref{12} correspond to the following set of the spin $1/2$
eigenfunctions for $H_D^{(+)}$:
\begin{gather*}
\Psi_0^{(+)}=\frac1{\sqrt{1+w\bar w}},
\qquad
\Psi_+^{(+)}=\frac{\bar w}{\sqrt{1+w\bar w}}.
\end{gather*}

The same similarity transformation transforms $H_D^{(-)} = /\!\!\!\!{\cal D}^2 \frac {1-\sigma_3}2 $ to the
Hamiltonian
\begin{gather*}%\label{rho02}
H_{(\rho=0)}'=-(1+w\bar w)^2\partial\bar\partial-(1+w\bar w)w\partial,
\end{gather*}
which coincides with the Hamiltonian of our model in the sector $F=2$ in the limit $\rho \to 0$! As was
discussed in Section~\ref{Se3}, it has the same spectrum as~\eqref{rho0} with the complex conjugated wave functions.
The corresponding realization of the ${\rm SU}(2)$ generators is as follows
\begin{gather}
J_+'=\partial+\bar w^2\bar\partial+\bar w,
\qquad
J_-'=\bar\partial+w^2\partial,
\qquad
J_3'=J-\frac12.
\label{su2prime}
\end{gather}

For completeness, it is instructive to explicitly give how the Hamiltonians $H_D^{(\pm)}$ look in our
notation,
\begin{gather*}
H_D^{(\pm)}
=-(1+w\bar w)^2\partial\bar\partial\pm\frac12(1+w\bar w)\left(w\partial_w-\bar w\partial_{\bar w}
\pm\frac12\right)+\frac14.
\end{gather*}
The relevant realizations of the ${\rm SU}(2)$ generators are obtained from~\eqref{Jpm} and~\eqref{su2prime} by
the same similarity transformation:
\begin{gather*}
J_+^{(\pm)}=\partial+\bar w^2\bar\partial\mp\frac12\bar w,
\qquad
J_-^{(\pm)}=\bar\partial+w^2\partial\pm\frac12w,
\qquad
J_3^{(\pm)}=J\pm\frac12.
\end{gather*}

To understand better what happens, let us look at the supercharges.
In the limit $\mu, \rho \to 0$, the supercharge $Q$ in~\eqref{Qquan} depends only on {\it one} holomorphic
fermion variable $\chi = (\psi_1 + i\psi_2)/\sqrt{2}$.
Similarly, $\bar Q$ depends only on $\bar \chi$:
\begin{gather}\label{reducedQ}
Q=-i\sqrt{2}\chi(1+w\bar w)\bar\partial,
\qquad
\bar Q=-i\sqrt{2}\bar\chi(1+w\bar w)\partial.
\qquad
\end{gather}
The similarity transformation
\begin{gather*}%\label{simQQbar}
(Q,\bar Q)\Rightarrow\sqrt{1+w\bar w}(Q,\bar Q)\frac{1}{\sqrt{1+w\bar w}}=(\tilde{Q},\bar{\tilde{Q}})
\end{gather*}
({\it the same} for $Q$ and $\bar Q$!) gives the supercharges
\begin{gather}\label{QDirac}
\tilde{Q}=-i\sqrt{2}\chi(1\!+\!w\bar w)\left[\bar\partial\!-\!\frac{w}{2(1\!+\!w\bar w)}\right],
\qquad
\bar{\tilde{Q}}=-i\sqrt{2}\bar\chi(1\!+\!w\bar w)\left[\partial\!-\!\frac{\bar w}{2(1\!+\!w\bar w)}\right].
\end{gather}
There is only one holomorphic fermion variable here, and the system can be described in terms of the $({\bf
2}, {\bf 2}, {\bf 0})$ superf\/ield.
The supercharges~\eqref{QDirac} coincide (if identifying $w \equiv \bar z$) with those in equation (3.26)
of~\cite{ISDir} brought on the sphere, with setting $W=0$ (no gauge f\/ield).
After mapping $\psi_{1,2} \to \sigma_{1,2}/\sqrt{2}$, the supercharges~\eqref{QDirac} are mapped onto
$/\!\!\!\!{\cal D} \frac {1\pm \sigma_3}2$.

Coming back to the spectrum of the full supersymmetric Hamiltonian in the limit $\rho\to 0$, it represents
{\it two copies} of the spectrum of the system~\eqref{reducedQ}.
Indeed, the full Hilbert space involves the functions $\Psi(w, \bar w, \chi, \lambda)$ where $\lambda =
(\psi_1 - i\psi_2)/\sqrt{2}$ is the holomorphic fermion variable orthogonal to $\chi$.
Thus, one can, e.g., multiply the fermion state of the system~\eqref{reducedQ} by $\lambda$ to obtain the
state in the sector $F=2$ of the full Hamiltonian with the same energy.

Note f\/inally that such a~nice interpretation reducing the problem to something already known is possible
{\it only} in the limit $\rho \to 0$.
When $\rho \neq 0$, the spectrum of the states with $J \neq 0$ does not coincide with~\eqref{specJ0} (this
is best seen if assuming $\rho$ in equation~\eqref{ep1} to be small and f\/inding out that the perturbative
corrections $\propto \rho$ to the spectrum~\eqref{specJ0} do not vanish) and we obtain something new.

\subsection*{Acknowledgements}

We are indebted to S.~Fedoruk for useful discussions.
E.I.\ would like to thank SUBATECH, Universit\'{e} de Nantes, for the warm hospitality in the course of
this study.
His work was carried out under the Convention N${}^{\rm o}$ 2010 11780.
He also acknowledges support from the RFBR grants 09-01-93107, 11-02-90445, 12-02-00517 and a~grant of the
IN2P3-JINR Programme for 2012.

\pdfbookmark[1]{References}{ref}
\LastPageEnding

\end{document}